 \documentstyle[aps,epsf,amssymb]{revtex}           

\begin{document}



\title{Non-Boltzmann classical correction to the velocity auto-correlation function 
for isotropic scattering in two dimensions}                          

\author{Alexander Dmitriev$^{1,2}$, Michel Dyakonov,$^1$ and R\'emi 
Jullien$^3$} 

\address{$^1$Laboratoire de Physique Math\'ematique$^\dagger$,  
Universit\'e
         Montpellier 2,
         place E. Bataillon, 34095 Montpellier, France\\
         $^2$A. F. Ioffe Physico-Technical Institute, 194021 St. 
Petersburg,
         Russia\\
         $^3$Laboratoire des Verres$^\dagger$,  Universit\'e Montpellier 2,
         place E. Bataillon, 34095 Montpellier, France\\
$^\dagger$ Laboratoire associ\'e au Centre National de la Recherche
Scientifique (CNRS, France).}


\maketitle


\begin{abstract}

The classical correction to the velocity auto-correlation function of non-interacting 
particles due to memory effects, which are beyond the Boltzmann equation, is calculated 
both analytically and numerically for the case of isotropic scattering in two dimensions.

\end{abstract}


\vskip 0.2in

{\bf 1.Introduction}

\vskip 0.1in
Some fourty years ago a number of pioneering theoretical works appeared 
\cite{dorfman,sengers,kawasaki,weinstock,haines,JVL1,JVL2}, which were devoted to the 
calculation of the memory-effect (non-Markovian) corrections to the kinetic coefficients 
and to the velocity auto-correlation function, through which they can be expressed.  
It was shown that such quantities as the diffusion coefficient, the electrical conductivity,
 viscosity etc, can not be expanded in powers of density, as it was assumed previously.  
Later, it was also demonstrated that the velocity auto-correlation function 
(which we abreviate below as "correlation function") does not decay exponentially at large 
times, as the Boltzmann equation predicts, but rather contains a slow power-law tail 
\cite{alder,ernst,bruin}.

These important deviations from the conventional picture based on the Boltzmann equation 
are due to memory effects neglected in the Boltzmann approach.  For the case of 
non-interacting particles performing diffusion in a static random potential field these 
effects are due to returns of the particle to previously visited regions. In the 
two-dimensional case it was shown by Ernst and Wejland \cite{ernst} that, 
because of returns, for times $t$ much greater than the mean free flight time $\tau$, 
the correlation function has a negative tail decaying as $1/t^2$. Bruin \cite{bruin} 
has performed numerical calculations for scattering by hard disks, which showed that 
this asymptotic behavior for $t>>\tau$ may appear only at very long times, it is not 
reached even at $t\simeq 10\tau$.

Returns after a single collision (see Fig. 1a) are important at $t\ 
$\raise -1.2mm\hbox{$\buildrel<\over\sim$}$\  \tau$. For $t<<\tau$ they give a contribution to
 the probability of return increasing as $1/t$. This, in turn, leads to a non-analytical 
in the small parameter $d/\ell=Nd^2$ correction to the diffusion coefficient on the order 
of $(d/\ell)\ln (\ell/d)$, where $d$ is the effective scattering diameter, 
$\ell = (Nd)^{-1}$ is the mean free path, and $N$ is the two-dimensional concentration 
of scatterers \cite{sengers,haines,peierls,hauge}. Returns after two or more collisions 
(see Fig. 1b) give a smaller correction on the order of $d/\ell$.

\begin{figure}
\epsfxsize=200pt{\epsffile{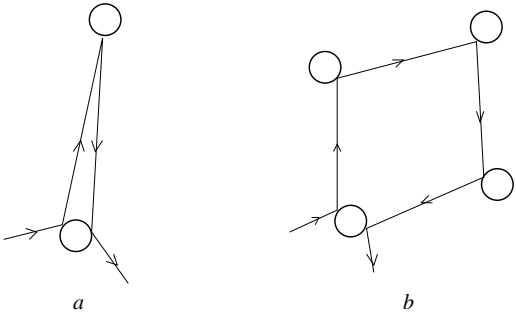}}

\vskip 15pt

\caption{ Illustration of the memory effect due to return to the same scattering center
after a single collision ($a$) or several collisions ($b$)}

\end{figure}

In our previous work \cite{dmitriev} we have drawn attention to another memory effect 
(the corridor effect) important in backscattering events. If a particle travels a distance 
$x$ after which it is backscaterred and returns to the initial point, the probability of 
this round trip of length $2x$ is proportional to 
$\exp(-x/\ell)$, not to  $\exp(-2x/\ell)$, as would suggest the conventional Boltzmann approach, 
since the existence of a free corridor of width $d$ allowing the first part of the journey 
garantees a collisionless return. 
This effect also gives a correction to the diffusion coefficient on the order of $d/\ell$. 

The role of memory affects becomes greatly enhanced in the presence of an applied magnetic 
field since the circling motion of an electron naturally increases the probability of returns, 
and many works were devoted to classical magnetoresistance, which for a degenerate 2D electron 
gas is entirely due to memory effects\cite{baskin}. In particular, it was shown that the 
irreversibility introduced by the magnetic field destroys the corridor effect resulting in an 
anomalous magnetoresistance in classically weak magnetic fields\cite{dmitriev}. 
An analytical theory of the anomalous low-field magnetoresistance was recently developped in Ref. 15.

The existing derivation \cite{JVL1,JVL2} of the memory effect corrections is rather long 
and cumbersome being based on a special technique of the so-called ring diagrams.

In the present paper we develop a relatively simple approach to the problem. We consider classical 
non-interacting particles with a fixed energy in two-dimensions scattered by randomly positioned 
centers with a given differential cross-section in the absence of magnetic field. We start 
with deriving a modified kinetic equation, which takes into account single returns after an 
arbitrary number of collisions. We then derive the leading logarithmic correction to the diffusion 
coefficient for an arbitrary angular dependence of the differential cross-section. Finally we find 
analytically the complete time dependence of the velocity correlation function for the special 
case of isotropic scattering taking into account the corridor effect and calculate the correction 
to the diffusion coefficient for this case. We perform numerical simulations for the cases of 
isotropic scattering and scattering by hard disks, and we find a good agreement between the 
numerical and analytical results.

\vskip 0.2in

{\bf 2. Derivation of the basic equation}
\vskip 0.1in

The derivation of the conventional Boltzmann equation from the Liouville
equation for a particle moving in the presence of randomly positioned scattering centers involves 
a number of simplifying assumptions. First, it is assumed that the mean free path, $\ell$, 
is much greater than the effective scattering diameter $d$ or, otherwise that $Nd^2<<1$, where $N$ 
is the concentration of scatterers. Second, it is assumed that the distribution function does not 
change much on the space scale on the order of $d$ and on the time scale on the order of $d/v$, 
where $v$ is the modulus of the particle velocity, which is conserved since the scattering is 
elastic. These two conditions make it possible to describe scattering in terms of the differential 
cross-section $\sigma (\phi)$ attributed to scattering centers positioned at given points in space. 
Third, memory effects are neglected: the Boltzmann approach is equivalent to randomly redistributing 
the scattering centers after each collision. Neglecting memory effects allows to take the average 
over the positions of scatterers in the Liouville equation, rather than use the average of its solution.

Restricting ourselves by the first two assumptions only, we can write down the following equation
 for the distribution function $f({\bf r},\phi,t)$ (see Appendix A and Ref. 15 for the derivation 
of this equation in the simplest case of scattering by hard disks).
$${\partial f \over \partial t}+{\bf v}{\partial f \over \partial {\bf r}}+v \sum_{i}
 {\delta({\bf r}-{\bf r}_{i})}\hat Tf=0,\eqno{(1)}$$
where ${\bf r}_{i}$ are the coordinates of the scattering centers, and $\hat T$ is the scattering 
operator proportional to the Boltzmann collision integral:
$$\hat Tf=\int_{0}^{2\pi}{\sigma (\phi-\phi')\bigl [f({\bf r},\phi, t)-f({\bf r},\phi', t)\bigr ]d\phi'},\eqno{(2)}$$

Note, that the distribution function, $f$, in Eq. (1) depends on the positions of scatterers, 
${\bf r}_{i}$, for a given realization. Also, since the scattering is elastic, the distribution 
function depends only on the polar angle, $\phi$, of the velocity vector {\bf v}. The conventional 
Boltzmann equation can be obtained from Eq. (1) by replacing the actual density of scatterers, 
$\sum_{i} {\delta({\bf r}-{\bf r}_{i})}$, by its average value, $N$. 

The correlation function, $K(t)$, (normalized by the condition $K(0)=1$) and the diffusion coefficient, 
$D$, can be expressed through the solution of Eq. (1) with the following initial and normalization conditions:
$$f({\bf r},\phi, 0)=\delta ({\bf r}) \delta (\phi),\qquad \int {f({\bf r},\phi, t)d{\bf r}d\phi} =1. \eqno{(3)}$$

Then
$$K(t)= \int {d{\bf r}} \int_{0}^{2\pi} {d\phi \cos \phi \langle f({\bf r},\phi, t)\rangle},  \eqno{(4)}$$
$$D={v^2 \over 2}\int_{0}^{\infty} {K(t)dt} . \eqno{(5)}$$

The angular brackets in Eq. (4) denote averaging over the positions  ${\bf r}_{i}$ of the scattering centers.

We now develop a method to calculate $K(t)$ and $D$ taking into account non-Boltzmann memory effects 
due to returns of the particle to previously visited scattering centers. As a first approximation 
we choose the solution $G({\bf r}-{\bf r'}, \phi, \phi', t)$ of the conventional Boltzmann equation:
$${\partial G \over \partial t} + {\bf v}{\partial G \over \partial {\bf r}}+vN\hat TG=0 \eqno{(6)}$$
with the initial condition 
$$G({\bf r}-{\bf r'}, \phi, \phi', 0)=\delta({\bf r}-{\bf r'})\delta(\phi -\phi'). $$

We re-write Eq. (1) in the form
$${\partial f \over \partial t}+{\bf v}{\partial f \over \partial {\bf r}}+Nv\hat Tf 
= -\nu ({\bf r})\hat Tf, \eqno{(7)}$$
where $\nu ({\bf r})=\sum_{i} {\delta({\bf r}-{\bf r}_{i})} - N$ is the fluctuation of the scatterer's 
concentration, and $\langle \nu ({\bf r})\rangle =0$. As above, the distrubution function $f$ 
in this equation depends on the actual positions of the scattering centers  ${\bf r}_{i}$.

Eq. (7) may be also written in an integral form using the formal solution of Eq. (6):
$$f({\bf r},\phi, t)=G({\bf r}, \phi, 0, t)-v\hat G\nu \hat Tf, \eqno{(8)}$$ 
where $\hat G$ is the integral operator with the kernel $G({\bf r}-{\bf r'}, \phi, \phi', t-t')$.

We now substitute $f$ given by Eq. (8) into the right-hand side of Eq. (7) and we take the average
 of the resulting equation over the positions of the scatterers ${\bf r}_{i}$. 
In doing this we must deal with the product of functions $\nu ({\bf r}), \nu ({\bf r'})$, 
 and the distribution function $f$, which all depend on the coordinates ${\bf r}_{i}$. 
 We decouple $f$ from the averaging procedure by writing
$$\langle \nu ({\bf r})\nu ({\bf r'})f \rangle \approx \langle 
\nu ({\bf r})\nu ({\bf r'})\rangle \langle f\rangle . $$

This approximation takes into account single returns to the same scattering center, but neglects multiple returns.

In the absence of correlation in the positions of the scatterers, which we assume to be true, we have 
$$\langle \nu ({\bf r})\nu ({\bf r'})\rangle =N \delta ({\bf r}-{\bf r'}).$$

Hence, we obtain the following equation for the averaged distribution function 
(to simplify the notations we replace hereafter $\langle f\rangle $ by $f$):
$${\partial f \over \partial t}+{\bf v}{\partial f \over \partial {\bf r}}+Nv\hat 
Tf =Nv^2 \int_{0}^{2\pi} {d\phi'} \int_{0}^{t} {dt' \hat TG(0,\phi'' -\phi',t-t')\hat Tf}. \eqno{(9)}$$
In the right-hand side of this equation the function $\hat Tf$ is a function of $\phi'$,  
the left operator $\hat T$ acts on a function of the variable $\phi''$.

The Green function $G(0,\phi'' -\phi',t)$ is the probability for a particle to return to the 
initial point after time $t$ with velocity directed at angle $\phi''$, provided that the initial 
velocity is directed at angle $\phi'$. Obviously, this probability should depend on the 
difference $\phi'' -\phi'$ only (this is not the case for $G({\bf r}-{\bf r'},\phi'', 
\phi',t)$ if ${\bf r}\neq {\bf r'}$ ).

The correction in the right-hand side of Eq. (9) coincides with the result of Weijland 
and  Van Leeuwen \cite{JVL2} obtained by the ring-diagram technique.

Eq. (9) is the basis for the following calculations. The solution of this equation is needed 
for calculating the correlation function $K(t)$ and the diffusion coefficient $D$ with the 
help of Eqs. (4), (5). In fact, because of the approximations made during our derivation 
the right-hand side of this equation should be considered as a small perturbation.

\vskip 0.2in

{\bf 3. Derivation of the correlation function and the diffusion coefficient}
\vskip 0.1in

To calculate $K(t)$ we need to know the integral of the distribution function over the coordinate:
$$F(\phi,t)=\int{f({\bf r}, \phi, t)d{\bf r}}. \eqno{(10)}$$

Integrating Eq. (9) over ${\bf r}$ we get
$${dF \over dt}+Nv\hat TF = Nv^2 \int_{0}^{2\pi}{d\phi'} \int_{0}^{t}{dt'
 \hat TG(0,\phi'' -\phi',t-t')\hat TF}. \eqno{(11)}$$

This equation can be further simplified by using the property that the functions 
$\exp(im\phi)$ are the eigenfunctions of the scattering operator $\hat T$:
$$Nv\hat T\exp (im\phi )=\gamma _{m}\exp (im\phi),$$
$$\gamma _{m}=Nv\int_{0}^{2\pi}{(1-\cos (m\phi))\sigma (\phi)d\phi }. \eqno{(12)}$$

Expanding the functions $F$ and $G$ in Eq. (11) in Fourrier series, we obtain
$${dF_{m} \over dt}+\gamma _{m}F_{m} = {2\pi \over N}\gamma _{m}^2
 \int_{0}^{t}{dt' G_{m}(0,t-t')F_{m}(t')}. \eqno{(13)}$$
where
$$F_{m}(t)={1 \over 2\pi }\int_{0}^{2\pi}{\exp(-im\phi)F(\phi,t)d\phi},$$
$$ G_{m}(0,t)={1 \over 2\pi }\int_{0}^{2\pi}{\exp (-im \phi )G(0,\phi ,t)d\phi }.$$

Eq. (13) allows to calculate any moment of the distribution function with the corrections
 due to returns taken into account.

As seen from Eqs. (4), (5), and (10), the correlation function and the diffusion coefficients
 can be expressed via the functions $F_{1}(t)$ and  $F_{-1}(t)=F_{1}^{*}(t)$:
$$K(t)=2\pi {\rm Re} F_{1}(t). \eqno{(14)}$$

Eq. (13) can be solved exactly. However, it should be noted that this equation, like our
 basic Eq. (9) takes into account single returns only and consequently, it gives the correct 
results only in the leading order in the small parameter $Nd^2$. For this reason, as we 
have already mentioned above, the right-hand term in Eq. (13) should be considered as a 
small perturbation. This means that in this term one can use the Boltzmann expression for $F_{m}(t)$:
$$F_{m}(t)= {1 \over 2\pi}\exp (-\gamma _{m}t), \eqno{(15)}$$
where the factor $(1/2\pi)$ appears because of the initial condition $F(\phi,0)=\delta(\phi)$.
 Substituting this expression onto the right-hand side of Eq. (13), for $m=1$ we get:
$${dF_{1} \over dt}+\gamma F_{1} = {\gamma ^2 \over N} \int_{0}^{t} 
{G_{1}(0,t-t')\exp (-\gamma t')dt'}, \eqno{(16)}$$
where 
$$\gamma \equiv \gamma _{1}=\tau ^{-1}=Nv\sigma_{tr}=Nv\int_{0}^{2\pi} 
{\sigma (\phi)(1-\cos \phi)d\phi}$$
is the inverse momentum relaxation time, or transport time, $\tau$.

The solution of Eq. (16) can be easily found. After some manipulation it can be presented as follows:
$$F_{1}(t)={1 \over 2\pi}\exp (-\gamma t)+{\gamma ^2 \over N}\int_{0}^{t} 
{G_{1}(0,t-t')\exp (-\gamma t') t' dt'}.  \eqno{(17)}$$

Finally, using Eq. (14) we obtain the following expression for the correlation function:
$$K(t)=K_{0}(t)+\delta K(t), \qquad \delta K(t)=2\pi {\gamma ^2 \over N}\int_{0}^{t}
{G_{1}(0,t-t')\exp (-\gamma t')t' dt'},  \eqno{(18)}$$
where $\delta K(t)$ is the correction to the Botzmann result, $K_{0}(t)=\exp (-\gamma t)$.

Substituting this expression into Eq. (5) and changing the order of integrations over 
$t$ and $t'$, we obtain a formula for the diffusion coefficient:
$$D=D_{0}+\delta D, \qquad D_{0}={v^2 \over 2\gamma }, \qquad \delta D = 
{\pi v^2 \over N}\int_{0}^{\infty}{G_{1}(0,t)dt}. \eqno{(19)}$$

Here $D_{0}$ represents the Boltzmann result, while $\delta D$ gives the correction due to returns, 
which is beyond the Boltzmann equation. One can see that calculation of these corrections 
is reduced to finding $G_{1}(0,t)$, which is equivalent to finding  $G(0,\phi ,t)$, 
the Boltzmann probability of return to the initial point ${\bf r}=0$ with velocity directed 
at angle $\phi$ with respect to the initial velocity.

\vskip 0.2in

{\bf 4. Contribution of returns after a single collision}
\vskip 0.1in

We have mentioned in the Introduction that the leading (logarithmic) correction to the diffusion 
coefficient is due to returns after a single collision (see Fig. 1). We will now show that 
the contribution of this process can be easily found for an arbitrary angular dependence of 
the scattering cross-section $\sigma (\phi)$. To do this, we must find the corresponding 
contribution to the function  $G(0,\phi ,t)$. 

We re-write Eq. (6) in an integral form:
$$G({\bf r},\phi ,t)=\delta ({\bf r}-{\bf v}_{0}t)\delta (\phi)\exp(-\gamma_{0}t)$$ 
$$+Nv\int_{0}^{2\pi}d\phi' \sigma(\phi -\phi') \int_{0}^{t}\exp 
(-\gamma_{0}(t-t'))G({\bf r}-{\bf v'}(t-t'),\phi' ,t')dt', \eqno{(20)}$$
where $\gamma_{0}=Nv\sigma$, $\sigma$ is the total scattering cross-section, 
${\bf v}_{0}$ is the initial velocity (with $\phi =0$), and $\phi'$ is the angle of the vector  ${\bf v'}$. 

Eq. (20) is convinient for obtaining the contribution of a given number of collisions 
before return by iterations. Since we are now interested in returns after a single collision, 
we can insert into the right-hand side of Eq, (20) the zero approximation 
for $G$, equal to $\delta ({\bf r}-{\bf v}_{0}t)\delta (\phi)\exp(-\gamma_{0}t)$. 
Denoting the correction thus obtained as $\delta G$, we get:
$$\delta G({\bf r},\phi ,t)=Nv\int_{0}^{2\pi}d\phi' \sigma(\phi -\phi')\int_{0}^{t}
\exp [-\gamma_{0}(t-t') -\gamma_{0}t']\delta ({\bf r}-{\bf v'}(t-t')-{\bf v}_{0}t')dt'.\eqno{(21)}$$

Putting ${\bf r}=0$ and taking into account the relation
$$\delta ({\bf v'}(t-t')+{\bf v}_{0}t')={\delta (\phi' -\pi) \over v^2(t-t')}\delta (t-2t'),$$
we find
$$\delta G(0,\phi ,t)={N \over v}{\exp(-\gamma_{0}t) \over t}\delta (\phi -\pi)\sigma(\phi).  \eqno{(22)}$$

Thus, for $\delta G_{1}(0,t)$ we obtain:
$$\delta G_{1}(0,t)=-{N \over 2\pi v}{\exp(-\gamma_{0}t) \over t}\sigma(\pi). \eqno{(23)}$$

To find the correction $\delta D$ to the diffusion coefficient we must substitute this expression 
in Eq. (19). The integral in Eq. (19) diverges logarithmically, which is the result of ignoring 
the finite radius of the scattering center. To avoid the divergency, we will replace $t$ in the 
denominator of the expression Eq. (23) by $t+t_{0}$, where the cut-off time $t_{0}$ is 
on the order of $\sigma /v$. Such a regularization seems reasonable, since it makes the return 
probability to be finite at $t \rightarrow 0$, as it should be if the finite radius of the 
scattering center is taken into account, and since our entire theory is applicable for time 
scales larger than  $\sigma /v$.

Eq. (23) must be also corrected to include the corridor effect \cite{dmitriev,cheianov}, 
which we mentioned in the introduction and which is not taken into account by Eq. (9). 
This effect is relevant for backscattering events, when the particle follows practically 
the same path (in the opposite direction) after a collision. The probability to make this round 
trip without collisions during time $t$ should be $\exp(-\gamma_{0}t/2)$ rather than 
$\exp(-\gamma_{0}t)$ as Eq. (23) says, because once the path $1\rightarrow 2$ during the 
time $t/2$ is collisionless, we are sure to have no collisions on the return path $2\rightarrow 1$. 
This effect can be taken into account by a more accurate evaluation of Eq. (1), which 
(within the approximation of point-like scatterers) contains all memory effects, 
including the corridor effect, see Ref. 15 where this was done for scattering by hard disks. 
Here we will simply modify Eq. (23) "by hand" replacing $\exp(-\gamma_{0}t)$ by $\exp(-\gamma_{0}t/2)$.

Thus, we replace Eq. (23) by
 $$\delta G_{1}(0,t)=-{N \over 2\pi v}{\exp(-\gamma_{0}t/2) \over t+t_{0}}\sigma(\pi). \eqno{(23a)}$$

We then obtain the correction to the diffusion coefficient due to returns after a single collision:
$${\delta D \over D_{0}}=N\sigma_{tr}\sigma(\pi)\Bigl ( -\ln ({1 \over \gamma_{0} t_{0}})+
{\bf C}-\ln 2 \Bigr ), \eqno{(24)}$$
where $D_{0}=v^2/2\gamma$ is the Boltzmann value of the diffusion coefficient and ${\bf C}=0.577$ 
is the Euler constant. Since the exact numerical coefficient in the argument of the logarithm is 
unknown, the constant ${\bf C}-\ln 2$ could be safely discarded.  We prefer to keep it in order 
to compare Eq. (24) with the more general formula derived in Sect. 5 and to have a clear definition 
of the cutoff parameter $t_{0}$ when discussing numerical results in Sect. 6.

With regard to this formula it should be noted that the argument of the logarithm contains the 
constant $\gamma_{0}$, which is expressed through the total scattering cross-section. 
The latter is equal to infinity for any realistic scattering potential that does not drop 
to zero at a finite distance (as it is the case for scattering on hard disks when the total 
scattering diameter is of course equal to the disk diameter). The divergency of the total 
cross-section is due to very small scattering angles. Since we consider the scattering centers 
as points, any scattering event will deviate the particle from its path connecting two 
scattering centers, however small the scattering angle may be. If the finite radius of the 
scatterer is taken into account, then scattering angles less than $\sim d/\ell$ will not 
matter anymore. This means that the integral over $\phi$, which gives the total scattering cross-section, 
entering the definition of $\gamma_{0}$, should in fact be truncated to exclude scattering angles 
less that $~d/\ell$. Because of the logarithmic dependence of $\delta D$ on $\gamma_{0}$, 
the exact value, which should be attributed to this cut-off, is not very important.

Substituting Eq. (23a) into Eq. (18), we obtain the correction to the correlation function 
in the time interval $t_{0} << t << \gamma_{0}^{-1}$:
$$\delta K(t)=-N \sigma_{tr} \sigma (\pi ) \gamma t \ln (t/t_{0}). \eqno{(25)}$$

To conclude this section we note that returns after multiple collisions, which are important 
for times $t \ $\raise -1.2mm\hbox{$\buildrel>\over\sim$}$\ \tau$, give a correction 
to the diffusion coefficient on the order of $N\sigma_{tr}^2$. Thus, as one can see from 
Eq. (24) the logarithmic correction dominates if $\sigma (\pi)$ is not too small, i.e. in all 
cases when the backscattering is not strongly supressed.

\vskip 0.2in

{\bf 5. Correlation function for isotropic scattering}
\vskip 0.1in

For the special case of isotropic scattering $\sigma_{tr}=\sigma$, $\sigma(\pi)=\sigma /2\pi$, 
and $\gamma_{0}=\gamma = \tau ^{-1}=Nv\sigma$. This simplification allows to find analytically the 
Green function $G(0,\phi,t)$ for arbitrary times (that are larger than the cutoff time $t_{0}$). 

The details of the calculation are given in Appendix B. The function  $G_{1}(0,t)$, through 
which the corrections to the correlation function and the diffusion coefficient are expressed, has the form:
$$G_{1}(0,t)={\gamma^2 \over 4\pi ^2v^2}\Bigl (-{\exp (-x/2) 
\over x+x_{0}}+{1-\exp (-x) \over x}+2\exp(-x)\bigl ({\rm Ei} (x)-{\rm Ei} (2x)\bigr )\Bigr ), \eqno{(26)}$$
where we have introduced the notations $x=t/\tau$, $x_{0} = t_{0}/\tau <<1$, and $ {\rm Ei} (x)= 
-\int_{-x}^{\infty} dy \exp (-y)/y$ is the integral exponent. Since $ t_{0}\sim \sigma /v$ 
and $\tau = (N\sigma v)^{-1}$, the cutoff parameter can be presented as $x_{0}=\alpha N\sigma^2$, 
where $\alpha$ is an unknown constant on the order of unity.

Comparing Eqs. (26) and (23a), one can see that the first term in Eq. (26) represents the contribution 
from returns after a single collision with the cutoff at $t \rightarrow 0$ and the corridor effect 
taken into account as described in the previous section. The other terms in Eq. (26) correspond to 
returns after two or more collisions and do not contain any singularities.

A direct calculation using Eqs. (19) and (26) gives us the correction to the diffusion coefficient:
$${\delta D \over D_{0}}={N\sigma^2 \over 2\pi} \Bigl (-\ln ({1 \over x_{0}})+
{\bf C}-2\ln 2 \Bigr ). \eqno{(27)}$$
which differs from Eq. (24) describing the contribution of returns after a single collision 
by an additional term $(-\ln 2)$ in the brackets, which is the contribution of returns 
after multiple collisions. 

The correction to the correlation function, $\delta K(t)$ can now be calculated by using the Eqs. (26) and (18). A direct calculation gives the following, rather cumbersome, expression (here again $x=t/\tau$):
$$\delta K(t)=-{N\sigma^2 \over 2\pi}\kappa (x),$$
$$\kappa (x) = 3e^{-x}-2e^{-x/2}-1+2e^{-2x}\Bigl ({\rm Ei} (2x)-{\rm Ei}(x)\Bigr)
+(x+x_0)e^{-x}\Bigl ({\rm Ei}({x+x_{0} \over 2})-{\rm Ei}({x_{0} \over 2})\Bigr )$$
$$+e^{-x}\Bigl [(x-2)\Bigl ({\rm Ei}(x)-{\bf C}-\ln x \Bigr)+x+2(x-1)\ln 2\Bigr ].  \eqno{(28)}$$

\begin{figure}

\epsfxsize=200pt{\epsffile{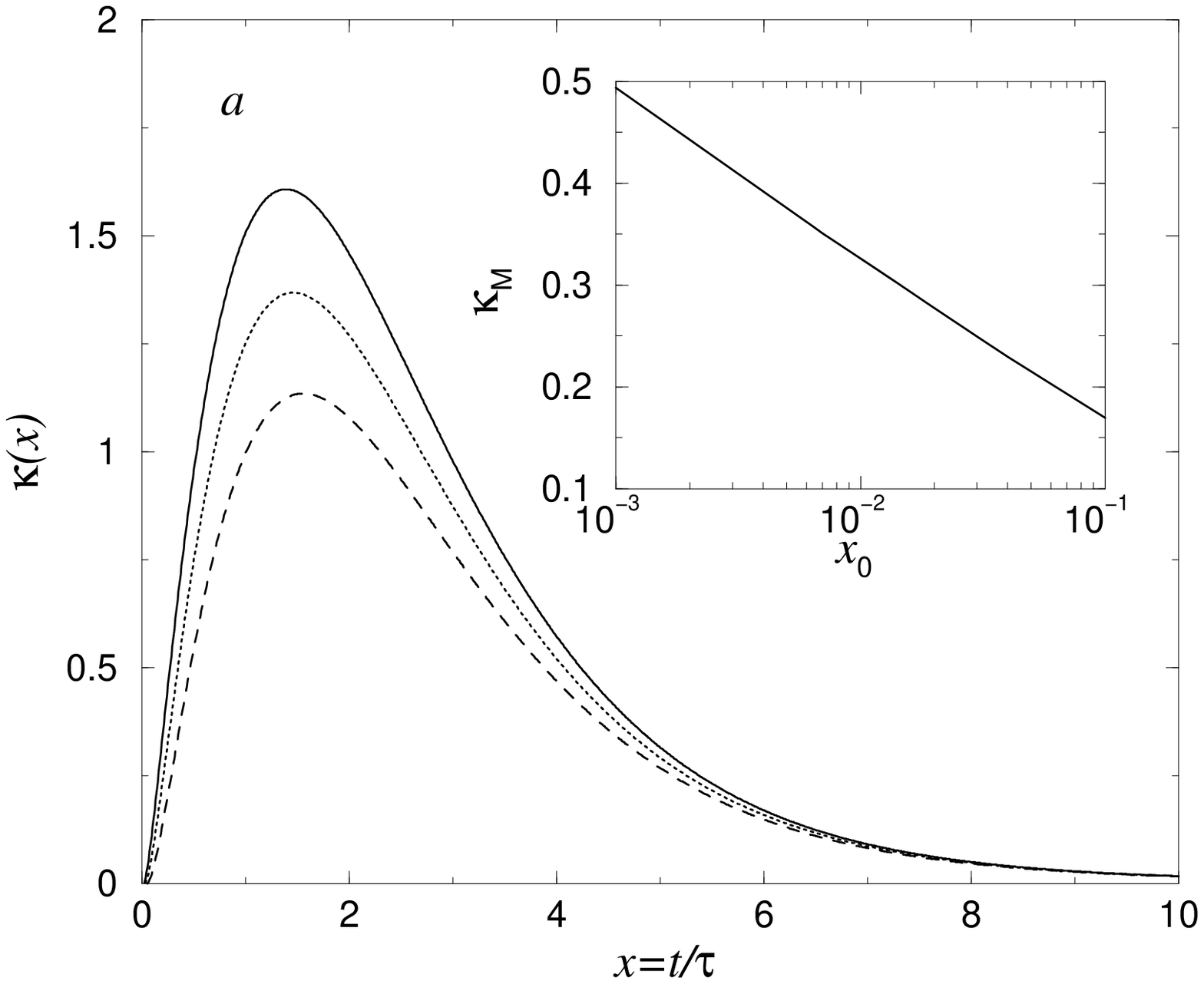}}
\epsfxsize=200pt{\epsffile{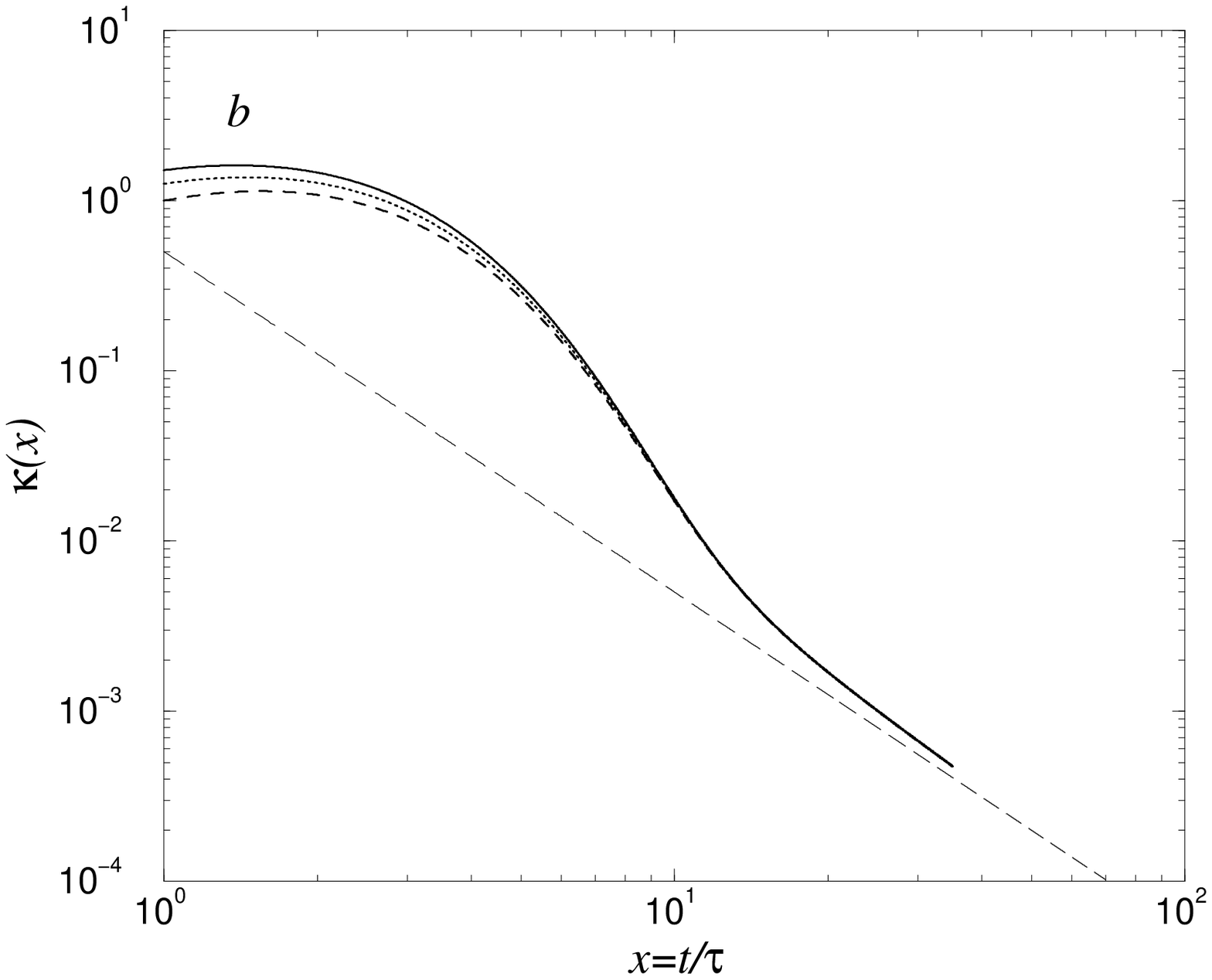}}

\caption{
a) Plot of $\kappa(x)=-(2\pi/N\sigma^2)\delta K(x)$
versus $x=t/\tau$ for different values of the cut-off parameter $x_0$.
Continuous, dotted and dashed curves correspond to $x_0$= 0.01, 0.02 and 0.04, respectively.
The inset shows the dependance of the maximum value, $\kappa_M$, on $\ln x_0$.
b) The long time part is emphasized in a log-log plot to show that the different
curves become quickly superimposed for $x>5$ and reach very slowly the
asymptotic regime, $1/2x^2$, as $x\rightarrow\infty$ (shown by the long-dashed line).
}
\end{figure}

The function $\kappa (x)$ is presented in Fig. 2a for several values of the cutoff parameter $x_{0}$.
In inset is shown the linear dependance of the maximum value of $\kappa(x)$ on $\ln x_0$.

For small times, $x_{0}<<x<<1$,  Eq. (28) gives
$$ \kappa (x) = x \ln (x/x_{0}), \eqno{(29)}$$ 
so that $\delta K(t)$ coincides with the expression given by Eq. (25), 
if one takes into account that for isotropic scattering  $\sigma_{tr}=\sigma$ 
and $\sigma(\pi)=\sigma /2\pi$.

At large times, $x>>1$, the correlation function is dominated by the contribution of returns over
 long diffusive trajectories involving many collisions. It can be presented as a series in inverse 
powers of $t$. Using Eq. (28) in this limit, we obtain the leading terms of this expansion:
$$\kappa (x)={1 \over 2x^2} \bigl (1+{5 \over x} \bigr ) . \eqno{(30)}$$

The leading term ($\sim 1/t^2$) coincides with the known result obtained by Ernst and Weyland, 
\cite{ernst} while the second one gives the first correction. 
We note that the leading term (but not the correction) is a universal result, which 
can be derived in the diffusion approximation \cite{ernst} and which does not depend on the 
scattering cross-section. It can be seen that the asymptotic $1/t^2$ behavior 
is approached very slowly, so that even for $t=50\tau $ ($x=50$) the correction still makes 
10\% (see Fig. 2b). This explains why the predicted asymptotic behavior was not 
reached in the numerical simulations done by Bruin \cite{bruin}, as well as in our 
simulations to be presented in the next section.

\begin{figure}

\epsfxsize=200pt{\epsffile{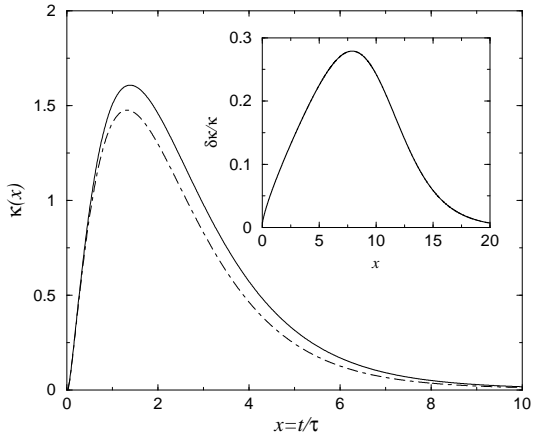}}

\caption{
Plot of $\kappa(x)$ (continuous line) and $\tilde\kappa(x)$ (dot-dashed line),
given by Eqs. (28) and (B8), respectively, for the same value of
the cut-off parameter, $x_0=0.01$. The relative difference $(\kappa-\tilde\kappa)/\kappa$ is 
shown in the inset.
}
\end{figure}

Finally, in Fig. 3, we present, for comparison, the function
$\tilde \kappa (x)$ given in Appendix B, Eq. (B8), in which the corridor effect is not
taken into account. One can see that the role of the corridor effect is noticable,
but rather small.

\vskip 0.2in

{\bf 6. Numerical simulations}
\vskip 0.1in

Using a  random number generator, the centers of $\cal N$ disks of diameter $d$ are uniformly randomly positionned on 
a plane inside a square box of edge length $L$. We take $L/d = 1000$ to be sure that $L$ remains more than one 
order of magnitude larger than the mean free path, even for the smallest concentrations that we have considered. 
The value of $\cal N$ is chosen to obtain the desired value for the dimensionless concentration, 
${\cal N} d^2 / L^2 = Nd^2$. Numerical simulations were performed for $Nd^2$ equal to 0.064, 0.032 and 0.016.
To calculate the velocity correlation function $K(t)$, as well as the diffusion coefficient $D$ given by Eq. (5),  
we first choose an initial point at random inside the box with an initial velocity direction arbitrarily chosen 
along the $x$-direction. We then determine the trajectory of a point-like particle  by joining the successive 
impact points,  an impact point being the first intersection, calculated analytically, of the linear trajectory 
with a disk periphery, the particle always coming from outside of the disk. We use standard numerical tricks to 
accelerate the search for impacts. When choosing the initial position, the disk interiors are not excluded 
(it was checked that excluding them introduces only a weak numerical difference,  which  vanishes in the limit 
$Nd^2\rightarrow 0$). The trajectory is made of successive straightline segments between collisions. 
Periodic boundary conditions are imposed at the edges of the square box. 

In the hard-disk (or Lorentz) model the scattering angle $\phi$ is related to the impact parameter $\rho$  
by $\rho = (d/2)\cos (\phi /2)$ (we consider $0<\phi <2\pi $, so that $\rho$ may be negative). The 
differential cross-section $\sigma (\phi )=|d\rho /d\phi |=(d/4)\sin (\phi /2)$ is anisotropic, with enhanced 
backscattering (in contrast to isotropic scattering by hard spheres in three-dimensions). As a consequence, 
the transport cross-section is {\it larger} than the total cross-section: $\sigma_{tr}=(4/3)d$.

To simulate isotropic scattering we define the scattering angle by the relation $\rho =(d/2)(1-\phi /\pi)$, 
so that $\sigma (\phi)=d/2\pi$. This relation is somewhat artificial, it leads to unrealistic situations, 
in which the disk may be cut by the scattering trajectory. This isotropic model, which is built to have an 
angle-independent differential scattering cross-section, does not correspond to any realistic potential and 
is considered for the sole purpose to check our analytical theory.

In order to calculate the correlation function, $K(x)$, and its dimensionless integral up to a time 
$t=x\tau$, $I(t) = \int_0^{t/\tau}K(x)dx = X(t)/v\tau$,
where $X(t)$ is the overal displacement of the particle in the $x$-direction during time $t$,  
we consider a set of 5000 discrete values $x_n$  of the reduced time $x=t/\tau$, regularly spaced 
up to $x_m =20$, and we calculate both quantities for each $x_n$. The values of $K(x_n)$ and $I(x_n)$ 
are averaged over 100 disk configurations and $10^8$  
trials for the starting point, except for the lowest concentration $Nd^2$=0.016 where only $10^7$ trials
for the starting point were considered. In this way we get practically continuous curves for both $K(t)$ and $I(t)$.
The reduced value of the diffusion coefficient, $D/D_0$, is obtained by extrapolation of the results for
$I(t)$ to $t \rightarrow \infty$. 

\begin{figure}

\epsfxsize=200pt{\epsffile{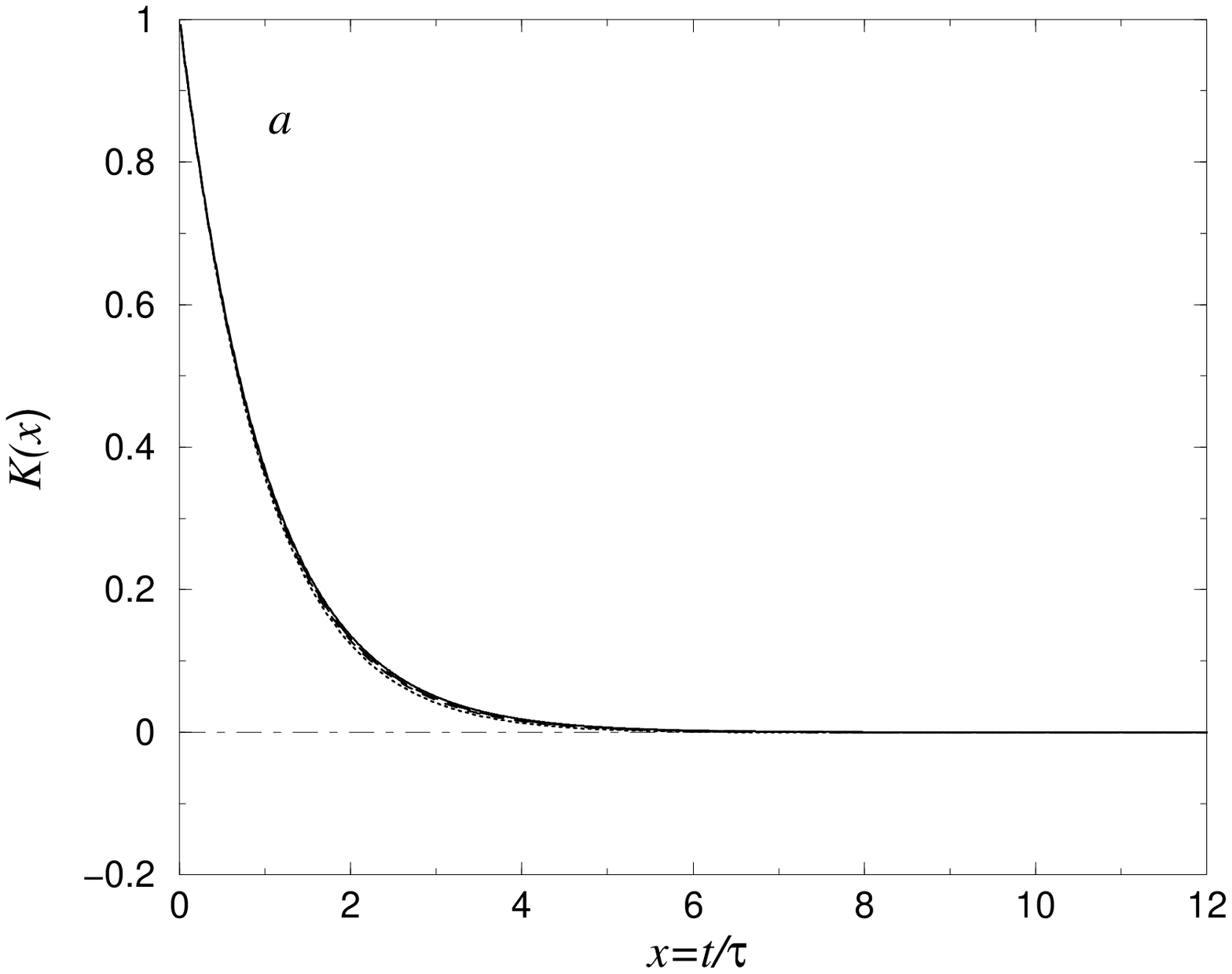}}

\epsfxsize=200pt{\epsffile{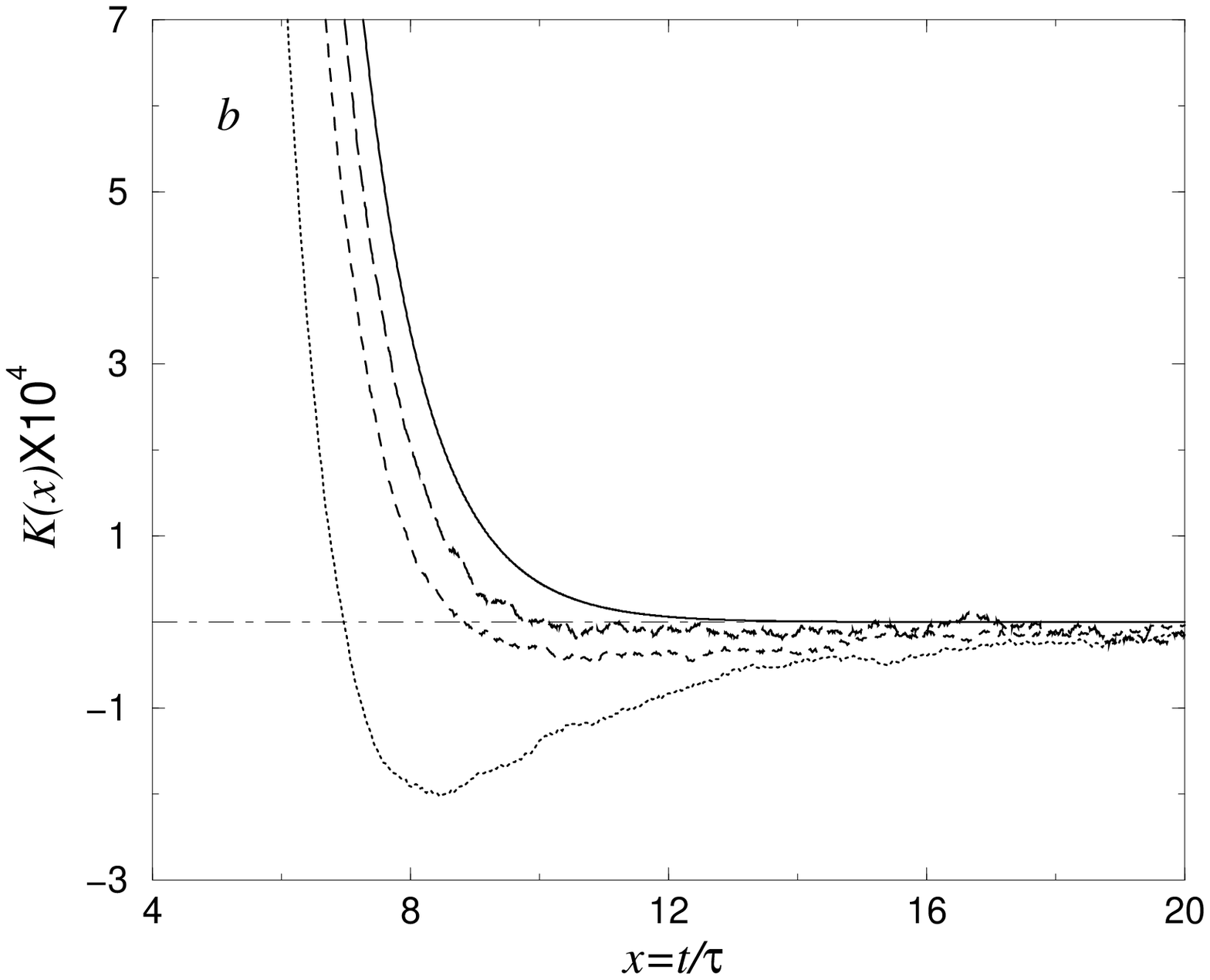}}

\caption{
a) Results of the numerical calculations for the isotropic model as a plot of $K(x)$
for different concentrations
Dotted, dashed and long-dashed curves correspond to $Nd^2$=0.064, 0.032 and 0.016,
respectively. The pure exponential $K_0=\exp(-x)$ is shown by the continuous curve.
b) For $t>4\tau$ the scale is enlarged by a factor $10^4$ to better show the
difference between $K(x)$ and $K_0(x)$.
}
\end{figure}

In Fig. 4, we present the numerical results for $K(t)$ for the isotropic scattering together 
with the pure exponential curve $K_0(t)=e^{-t/\tau}$ expected from the Boltzmann approach. 
The enlarged scale in Fig. 4b allows to clearly see the negative departure from the exponential 
behavior and the long-time tail of the correlation function. We have also calculated numerically
the correlation function for the hard disk scattering. The results are qualitatively similar
to the isotropic case and agree with the numerical simulations of Bruin\cite{bruin}.

Figure 5 presents the comparison of the numerical results for the normalized correction to the 
correlation function, $\kappa (x)$, for isotropic scattering with the theoretical formula given 
by Eq. (28). The only fitting parameter is the  the constant $\alpha$ in the definition of 
the cutoff value of $x_0 =\alpha Nd^2$. By choosing $\alpha =0.424$ we obtain a practically 
perfect fit to the theoretical curves for several values of $Nd^2$. The numerical errors are 
responsible for the noisy character of the simulation data for $x>10$, however they are 
practically negligible for smaller values of $x$, the error bar being smaller than the symbol size. 

\begin{figure}

\epsfxsize=200pt{\epsffile{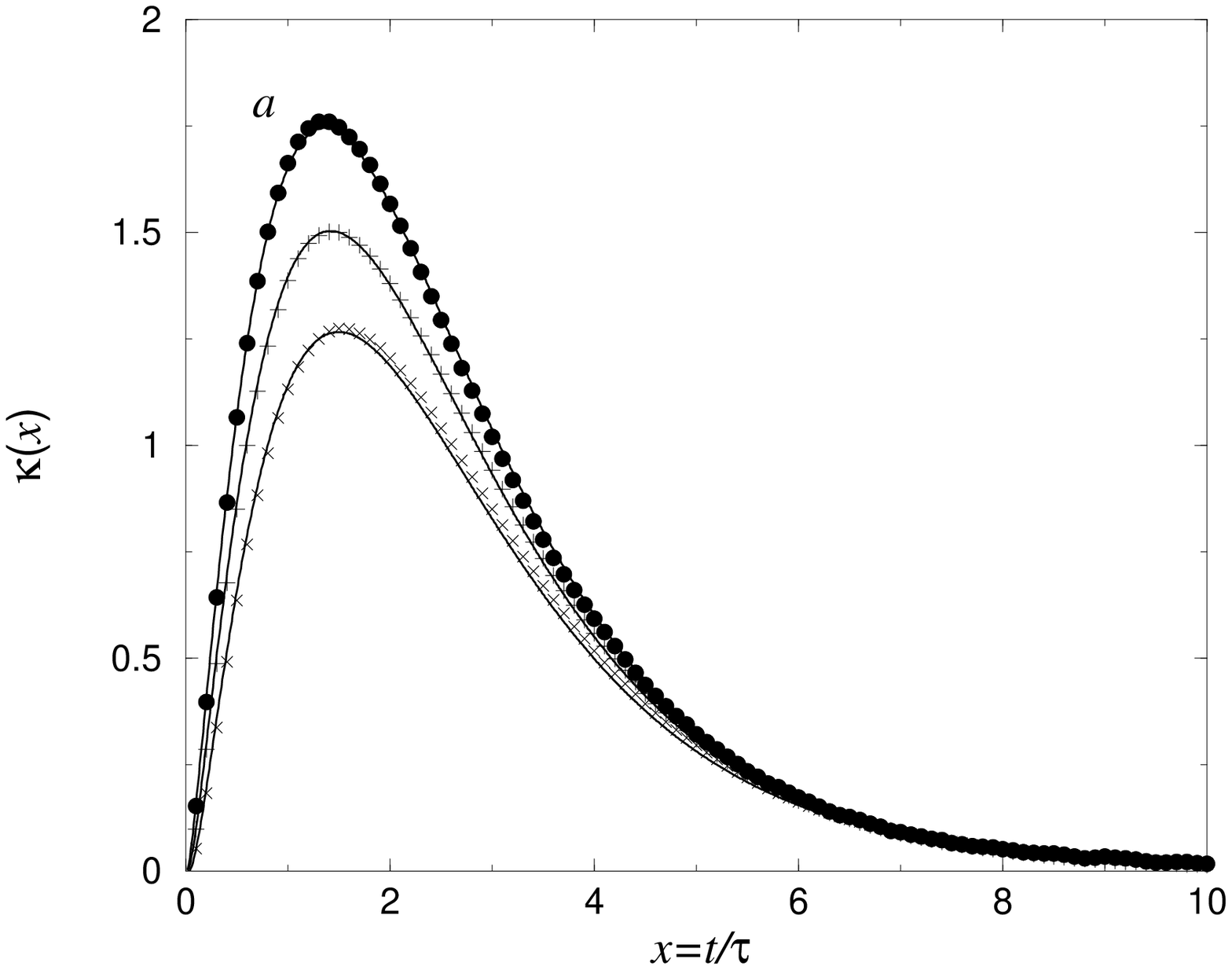}}

\epsfxsize=200pt{\epsffile{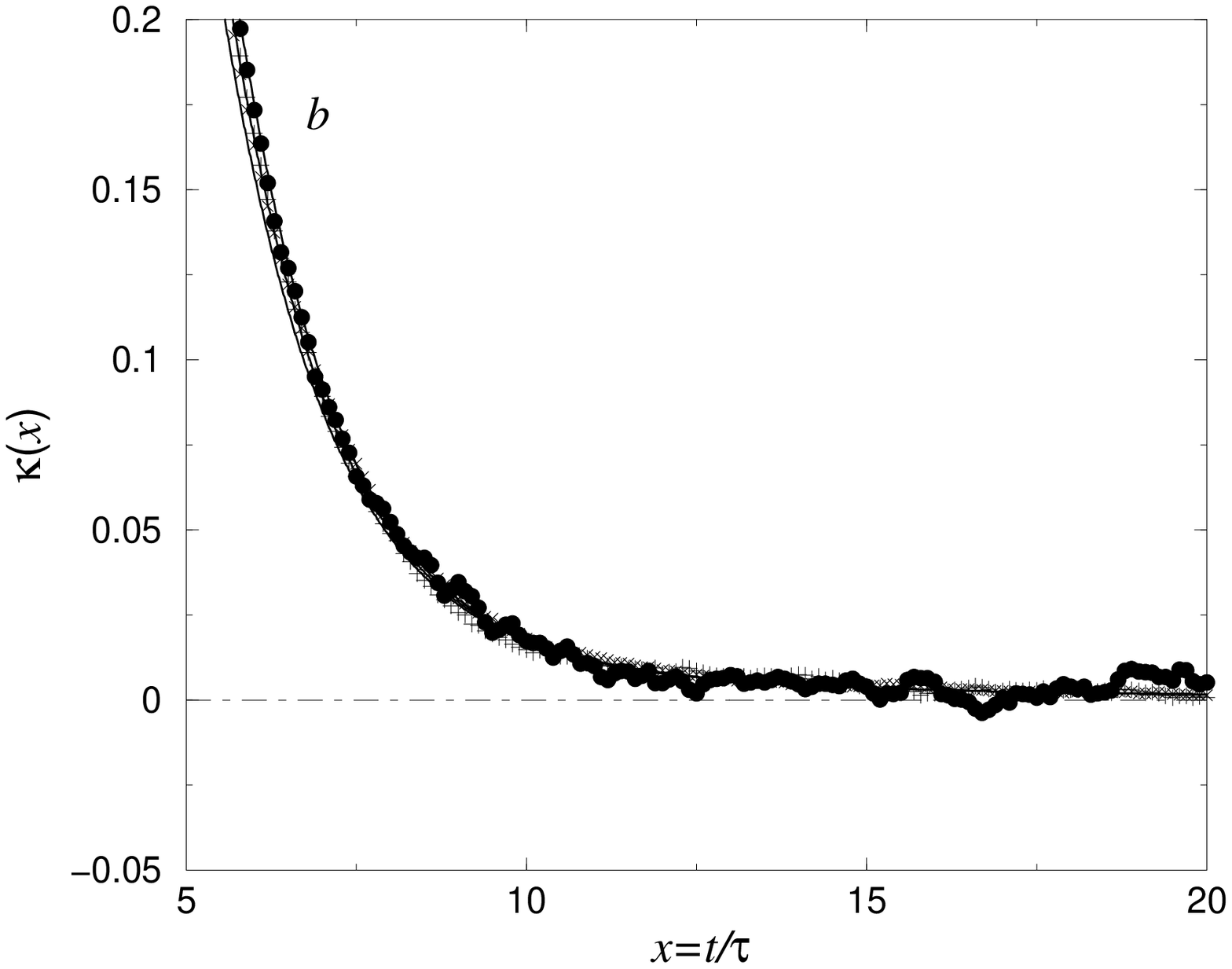}}

\caption{
a) The results of the preceeding figures are reported in a plot of $\kappa(x)$.
Only 200 points are selected in the range $0<x<20$. Crosses, plus signs
and filled circles correspond to $Nd^2$=0.064, 0.032 and 0.016,
respectively. The continuous lines correspond to the fits obtained using Eq. (28)
with the same proportionality factor. $\alpha=0.424$,  between $x_0$ and $Nd^2$ in the three cases.
The error bars are smaller than the symbol size.
b) The region $x>5$ is shown with a larger scale.                                         
At this new scale one can see the fluctuations of the numerical data giving
an idea of the magnitude of the error bars (which are
larger in the case $Nd^2=0.016$ due to the poorer statistics).
}
\end{figure}



Because of the high precision of our numerical simulation, we are able to verify the theoretical prediction 
concerning the role of the corridor effect. For this purpose we try to fit the numerical data to the 
function $\tilde \kappa (x)$ given by Eq. (B8), which does not take care of this effect, see Fig. 6. 
It is not possible to have a good fit with the same choice of the constant $\alpha$ as before. Thus we 
chose a different value, $\alpha = 0.283$, to
fit the maxima of the numerical curves. As it can be seen from Fig. 5, though the resulting fit could 
be considered as satisfactory, the difference is well beyond the numerical errors, in 
particular the numerical data points for $5<x<10$ are systematically above the theoretical curves. 
Comparing Figs. 5 and 6, one can see that taking the corridor effect into account makes the agreement 
between the theoretical and numerical curves substantially better.

Finally, in Fig. 7 we present the numerical results for the correction to the diffusion 
coefficient $\delta D/D_0$ in units of $Nd^2/2\pi $ versus $\ln (1/Nd^2)$, see Eq. (27). 
Data for both the hard-disk scattering and isotropic scattering are presented for several 
values of $Nd^2$.  For the case of isotropic scattering we get an excellent agreement with Eq. (27), 
in which the cutoff parameter is chosen as in Fig. 4: $x_{0}=0.424Nd^2$. The slope of the 
dependence on $\ln (1/Nd^2)$ in the general case is given by Eq. (24).  
For the hard-disk scattering the slope is equal to $1.95\pm 0.10$, 
which is close to what follows from Eq. (24). Indeed, for a given diameter $d$,  
the ratio of $\sigma_{tr}\sigma (\pi)$ for the two cases is equal to $2\pi /3=2.094$.

\begin{figure}
\epsfxsize=200pt{\epsffile{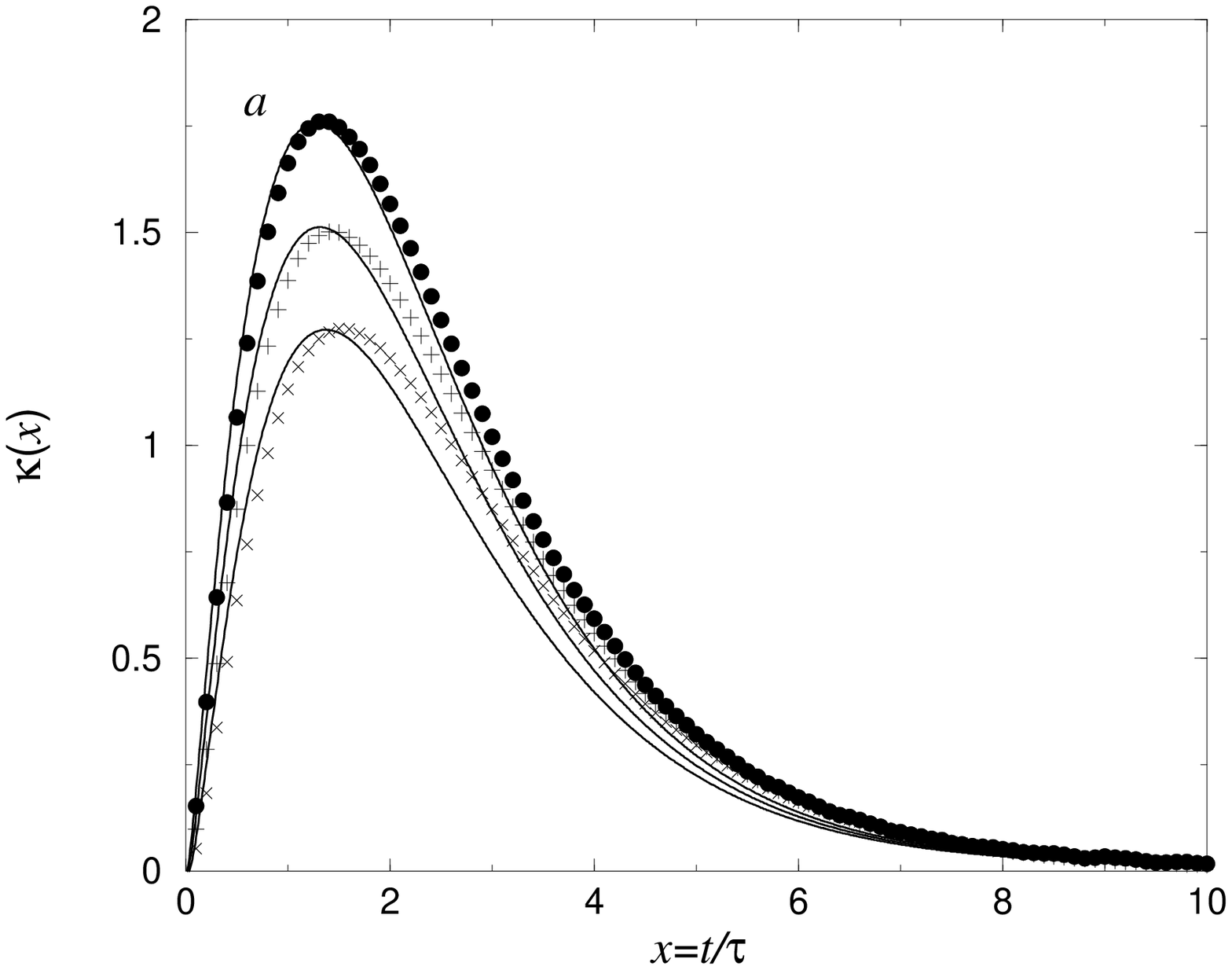}}

\epsfxsize=200pt{\epsffile{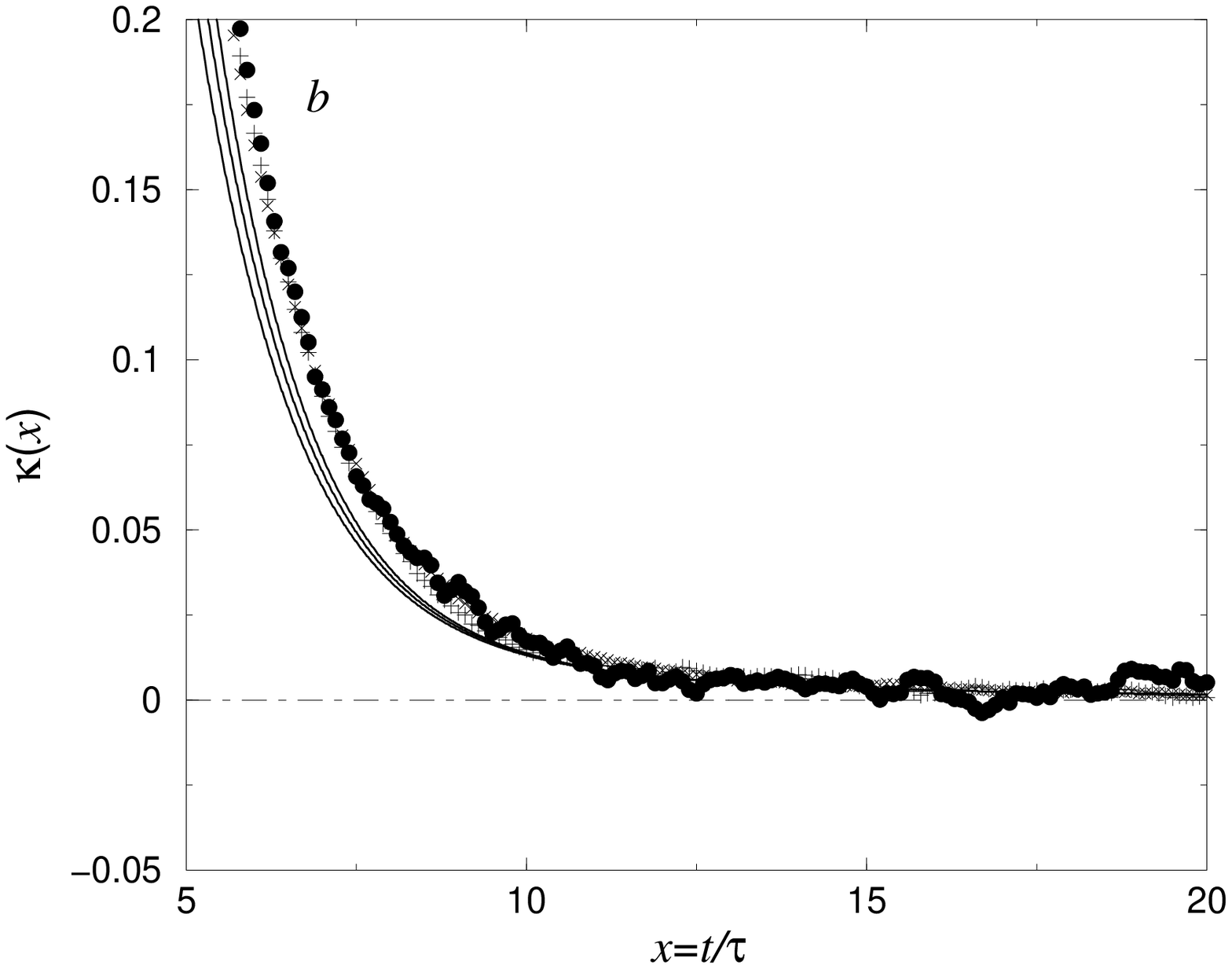}}

\caption{
a) The same numerical results are fitted by Eq. (B8),
which does not take care
of the corridor effect. Here, to fit the maximum values, we take
a different proportionality factor than in figure 5, namely
$\alpha=0.283$.
b) The region $x>5$ is shown with a larger scale.
}
\end{figure}

\begin{figure}

\epsfxsize=200pt{\epsffile{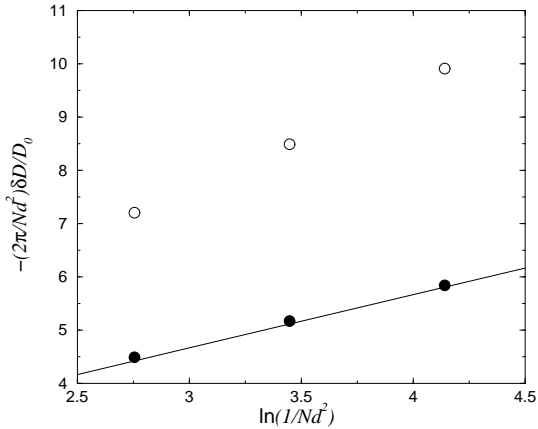}}

\caption{
Numerical results for the diffusion coefficient as a plot of $-(2\pi/Nd^2)\delta D/D_0$
versus $\ln(1/Nd^2)$ for the hard disk scattering (open circles) and for isotropic scattering
(filled circles). The linear behavior predicted by formula (27) in the isotropic case,
with $\alpha = 0.424$ as in Fig. 5, is represented
by the straight line (of slope 1). A linear fit through the three points in the hard disk case gives a
slope of $1.95\pm0.10$ in good agreement with the theoretical prediction of 2.094.
}
\end{figure}

\vskip 0.2in
{\bf 7. Conclusion}
\vskip 0.1 in

In this paper, we have presented a simple approach to the problem of classical corrections to the 
Boltzmann equation, which are 
due to memory effects. We have performed both analytical and numerical calculations of the 
velocity autocorrelation function for
non-interacting particles scattered by randomly located centers in two dimensions. 

In the particular case of isotropic scattering, we were able to provide a full analytical 
expression for the correction to the 
Boltzmann result, due to returns of the particles to previously visited regions, 
in excellent agreement with the results of numerical
simulations. Furthermore, the comparison between analytical and numerical results 
demonstrates that it is essential to take
care of the "corridor effect" associated with backscattering events. 
In the case of hard disk scattering (Lorentz model), we obtain
a good agreement for the dependence of the diffusion coefficient on the concentration of scattering centers.

A full theory taking into account both classical and quantum (weak localization) 
memory effects is needed to understand the relative
role of classical and quantum corrections depending on the ratio of the 
De Broglie wavelength to the scattering diameter.

\vskip 0.2in

{\bf Appendix A}
\vskip 0.1in

For the case of scattering by hard disks, the Liouville equation has the form \cite{JVL1,JVL2,cheianov}:
$${\partial f \over \partial t}+{\bf v}{\partial f \over \partial {\bf r}}+v 
(\hat T_{-}+\hat T_{+})f=0,\eqno{(A1)}$$
where the operators $\hat T_{-}$ and $\hat T_{+}$ describe scattering by the disks. 
They are defined by the equations:
$$\hat T_{-}f({\bf r},\phi , t)=\sum_{i}\int_{0}^{2\pi} d\phi' \delta 
(({\bf r}-{\bf r}_{i}-{\bf d}/2)\sigma(\phi -\phi')f({\bf r},\phi , t),\eqno{(A2)}$$

$$\hat T_{+}f({\bf r},\phi , t)=-\sum_{i}\int_{0}^{2\pi} d\phi' \delta 
(({\bf r}-{\bf r}_{i}+{\bf d}/2)\sigma(\phi -\phi')f({\bf r},\phi' , t).\eqno{(A3)}$$

Here $\sigma(\phi)=(d/4)\sin (\phi /2)$ is the differential cross-section for a disk of 
diameter $d$, the vector ${\bf d}$ is expressed through the velocities before and after 
the collision, ${\bf v}$ and ${\bf v'}$ by the relation (see Fig. 8):
$${\bf d}=d{{\bf v'}-{\bf v} \over \sqrt {2(v^2-{\bf v}{\bf v'})}}.\eqno{(A4)}$$

Equation (A1) is exact, the operators  $\hat T_{-}$ and $\hat T_{+}$ describing the mechanics of 
collisions (assuming that the disks do not overlap). The term $\hat T_{-}f$ describes the outflux of 
particles from the state with velocity ${\bf v}$ at the point ${\bf r}_{i}+{\bf d}/2$ on the 
disk edge, while the term  $\hat T_{+}f$ describes the influx of particles to the state with 
velocity  ${\bf v}$ at the point ${\bf r}_{i}-{\bf d}/2$ (see Fig. 8).

\begin{figure}

\epsfxsize=200pt{\epsffile{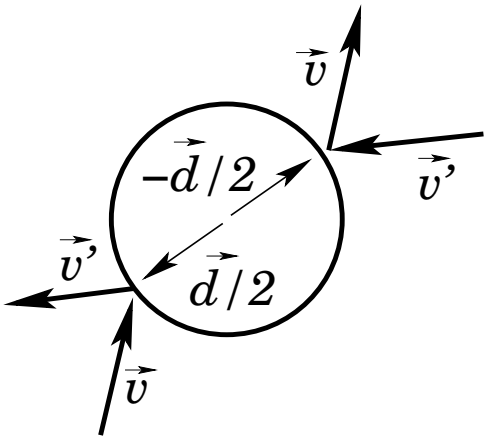}}

\caption{Scattering of a particle by a hard disk. Scattering events described by the
operators $T_-$ and $T_+$ are shown.
}
\end{figure}

If we make the assumption that the distribution function $f({\bf r},\phi , t)$ can be regarded
 as constant on distances on the order of $d$, we can neglect the vector ${\bf d}/2$ in the arguments 
of the $\delta$-functions in Eqs. (A2) and (A3). Then we immediately obtain Eq. (1).

\vskip 0.2in

{\bf Appendix B}
\vskip 0.1in

For the case of isotropic scattering, we write the the Boltzmann equation for the function $G$ 
in the form
$${\partial G \over \partial t}+{\bf v}{\partial G \over \partial {\bf r}}+\gamma (G-G_{0})=
\delta ({\bf r})\delta (\phi)\delta (t), \eqno{(B1)}$$
where
$$G_{0}({\bf r},t)={1 \over 2\pi}\int_{0}^{2\pi} G({\bf r}, \phi , t) d\phi$$
is the isotropic part of $ G({\bf r}, \phi , t)$.

Performing the Fourier transformation of this equation in variables ${\bf r}$ and $t$, we obtain:
$$-i(\omega-{\bf k}{\bf v}) G({\bf k}, \phi , \omega)+\gamma \bigl ( G({\bf k}, \phi , \omega)
- G_{0}({\bf k}, \omega) \bigr )=\delta (\phi). \eqno{(B2)}$$

If we find $G({\bf k}, \phi , \omega)$ from this equation and take its average over the angle 
$\phi$, we will have an algebraic equation for $ G_{0}({\bf k}, \omega)$. Solving it and 
using Eq. (B2), we obtain the following expression for the function $G({\bf k}, \phi , \omega)$:

$$G({\bf k}, \phi , \omega)={\delta (\phi) \over \gamma -i\omega +ikv\cos \phi_{k}}+$$
$$+{\gamma g(k,\omega) \over 2\pi (\gamma -i\omega +ikv\cos (\phi -\phi_{k}))
(\gamma -i\omega +ikv\cos \phi_{k})(g(k,\omega )-\gamma )}, \eqno{(B3)}$$
where $\phi_{k}$ is the angle of the vector ${\bf k}$ and $ g(k,\omega)=
\sqrt {(\gamma -i\omega)^{2} +k^{2} v^{2}}$.

Now we can easily find the expression for $G_{1}({\bf k}, \omega )$:
$$G_{1}({\bf k}, \omega )={1 \over 2\pi (\gamma -i\omega +
ikv\cos \phi_{k})}+{\gamma \exp (-i\phi_{k}) \over 2\pi ikv}{ g(k,\omega)-
\gamma +i\omega \over (\gamma -i\omega +ikv\cos \phi_{k})(g(k,\omega)-\gamma )}.  \eqno{(B4)}$$

The function $G_{1}(0,t)$ entering Eqs. (18) and (19) is given by the Fourier transform of 
Eq. (B4) for ${\bf r}=0$. The first term in Eq. (B4) then becomes $\exp (-\gamma t)\delta ({\bf v}_{0}t)$. 
This contribution should be ommited, since it does not describe returns for $t>0$. 
We begin the Fourier transformation of the second term by calculating the integral over 
the angle $\phi_{k}$ using the formula \cite{ryzhik}
$$\int_{-\pi}^{\pi}{exp(-i\phi) \over 1+a\cos \phi}d\phi=2\int_{0}^{\pi}
{\cos \phi \over 1+a\cos \phi }d\phi ={2\pi \over a}{\sqrt {1-a^2} -1 \over \sqrt {1-a^2}}. $$

Then we have
$$G_{1}(0,t)=-{\gamma \over (2\pi )^2 v^2}\int_{-\infty}^{\infty}{d\omega \over 2\pi }
e^{-i\omega t}\int_{0}^{\infty}{dk \over k}{(g(k,\omega )-\gamma +i\omega )^2 \over  g(k,\omega )
(g(k,\omega )-\gamma )}. \eqno{(B5)}$$

Instead of $k$ we now introduce a new variable $z=g(k,\omega )-\gamma + i\omega $, which gives us
$$G_{1}(0,t)=-{\gamma \over (2\pi )^2 v^2}\int_{-\infty}^{\infty}{d\omega \over 2\pi }
e^{-i\omega t}\int_{0}^{\infty }{zdz \over (z+2(\gamma -i\omega ))(z-i\omega)}. \eqno{(B6)}$$

The integral over $\omega$ can be now easily calculated and the remaining integral over $z$ gives:
$$G_{1}(0,t)={\gamma^2 \over 4\pi ^2v^2}\Bigl (-{\exp (-x) \over x}+{1-\exp (-x) \over x}
+2\exp(-x)\bigl ({\rm Ei} (x)-{\rm Ei} (2x)\bigr )\Bigr ), \eqno{(B7)}$$
where $x=\gamma t$.
After replacing $\exp (-x)$ by $\exp (-x/2)$ in the first term to account for the corridor 
effect and introducing the cutoff at $t_{0}=x_{0}/\gamma$ as described in Sect. 4, we obtain Eq. (26).

If we neglect the corridor effect (but still introduce the cutoff at $t_{0}$), Eq. (B7) can be 
used directly to calculate the correction to the correlation function. Inserting Eq. (B7) into Eq. (18) 
we get, instead of $\kappa (x)$ given by Eq. (28), the following expression for 
$\tilde \kappa (x)$, which does not take into account the corridor effect:
$$\tilde \kappa (x) = -1+2e^{-2x}\Bigl ({\rm Ei} (2x)-{\rm Ei}(x)\Bigr)
+e^{-x}\Bigl [1+(x-2)\Bigl ({\rm Ei}(x)-{\bf C}-\ln x \Bigr)+2(x-1)\ln 2
+(x+x_0)\ln (1+{x \over x_{0}})\Bigr ].  \eqno{(B8)}$$

\newpage


\newpage

\end{document}